\documentclass[aps, a4paper,superscriptaddress, nofootinbib, showpacs, 12pt]{revtex4}
\usepackage{amsmath,amssymb, bm, natbib}
\usepackage{dsfont}
\usepackage{epsfig}
\usepackage{graphicx}
\usepackage{slashed}
\usepackage{color}

\usepackage{accents}
\usepackage{mathrsfs}
\usepackage[center,footnotesize,hang]{subfigure}
\usepackage{bigstrut}

\def\br{\begin{array}}
\def\er{\end{array}}
\def\Dmott{\Delta m^2_{21}}
\def\Dmttht{\Delta m^2_{32}}

\def\D{\mathrm{d}}

\begin{document}

\title{Predictions from High Scale Mixing Unification Hypothesis}
\author{Gauhar Abbas}
\email{gauhar@imsc.res.in}
\affiliation{The Institute of Mathematical Sciences, Chennai 600 113, India}
\author{Saurabh Gupta}
\email{saurabh@imsc.res.in}
\affiliation{The Institute of Mathematical Sciences, Chennai 600 113, India}
\author{G. Rajasekaran}
\email{graj@imsc.res.in}
\affiliation{The Institute of Mathematical Sciences, Chennai 600 113, India}
\affiliation{Chennai Mathematical Institute, Siruseri 603 103, India}
\author{Rahul Srivastava}
\email{rahuls@imsc.res.in}
\affiliation{The Institute of Mathematical Sciences, Chennai 600 113, India}


\begin{abstract}

We investigate the renormalization group evolution of masses and mixing angles of Majorana neutrinos under the `High Scale Mixing Unification' hypothesis. Assuming the unification of quark-lepton mixing angles 
at a high scale, we show that all the experimentally observed neutrino oscillation parameters can be obtained, within 3-$\sigma $ range, through the running of corresponding renormalization group equations 
provided neutrinos have same CP parity and are quasi-degenerate. 
One of the novel results of our analysis is that $\theta_{23}$ turns out to be  non-maximal and lies  in the second octant. Furthermore, we derive new constraints on the allowed 
parameter space for the unification scale, SUSY breaking scale and $\tan \beta$, for which the `High Scale Mixing Unification' hypothesis works.

\end{abstract}


\pacs{14.60.Pq, 11.10.Hi, 11.30.Hv, 12.15.Lk}

\maketitle


\section{INTRODUCTION}
\label{sec1}


The quest for a unified theory of quarks and leptons is one of the  main goals of beyond standard model physics.  To this end, the unification of mixing angles of quarks and leptons, at a high scale, 
seems to be an exciting possibility. In the past, it has been investigated under the hypothesis referred to as `High Scale Mixing Unification' (HSMU) for the case of Majorana neutrinos 
\cite{Mohapatra:2003tw,Mohapatra:2004vw,Mohapatra:2005gs,Mohapatra:2005pw,Agarwalla:2006dj} and, recently, for the case of Dirac neutrinos \cite{Abbas:2013uqh}. 
A similar possibility has also been investigated in \cite{Haba:2012ar}. Within the HSMU hypothesis, the observed values of  oscillation parameters at low energies 
are obtained through the renormalization group (RG) evolution of these parameters from the unification scale (high scale) to the low scale $M_Z$ (mass of the Z boson). 

In addition, the HSMU hypothesis also provides hints about the type and nature of the underlying unified theory that might exist at the unification scale. One of the 
key predictions of the HSMU hypothesis is the small non-zero value of $\theta_{13}$ \cite{Mohapatra:2003tw,Mohapatra:2004vw,Mohapatra:2005gs,Mohapatra:2005pw,Agarwalla:2006dj}. At the time of the earlier 
work on HSMU, only an upper bound on $\theta_{13}$ existed and it was not known whether $\theta_{13}$ was zero or non-zero. 

The  recent results  from different experiments have established the value of $\theta_{13}$ to be non-zero 
\cite{Abe:2011sj,Adamson:2011qu,Abe:2012tg,Ahn:2012nd,An:2012eh}.  This precise measurement can be used to test predictions of various models and put stringent constraints on them. The current global 
scenario of the neutrino oscillation parameters (for normal hierarchy pattern)  \cite{Fogli:2012ua, GonzalezGarcia:2012sz} is summarized in the Table \ref{tab1}. Since $\theta_{13}$ is fairly well determined now, it is important 
to check whether HSMU is consistent with this measurement.

\begin{table}[h]\label{tab1}
\begin{center}
\begin{tabular}{|c|c|c|}
  \hline
  Quantity                                       & Best Fit $\pm 1$-$\sigma$                                   &  3-$\sigma$ Range\\
  \hline
  $\Delta m^2_{21}~(10^{-5}~{\rm eV}^2)$         & $7.50^{+0.18}_{-0.19}$                                    &  7.00 -- 8.09         \\
  $\Delta m^2_{31}~(10^{-3}~{\rm eV}^2)$         & $2.473^{+0.070}_{-0.067}$                                 &  2.276 -- 2.695        \\
  $\theta_{12} /^{\circ}$                        & $33.36^{+0.81}_{-0.78}$                                   &  31.09-- 35.89           \\
  $\theta_{23} /^{\circ}$                        & $40.0^{+2.1}_{-1.5} \oplus 50.4^{+1.3}_{-1.3}$            &  35.8 -- 54.8   \\
  $\theta_{13} /^{\circ}$                        & $8.66 ^{+0.44}_{-0.46}$                                   &  7.19 -- 9.96     \\
  \hline
  \end{tabular}
\end{center}
\caption{The global fits for neutrino oscillation parameters \cite{GonzalezGarcia:2012sz}.}
\end{table}    

Furthermore, with the operation of the Large Hadron Collider (LHC), two important developments have occurred. What is presumably the long awaited Higgs boson has been discovered with a mass around 125 GeV \cite{Aad:2012tfa,Chatrchyan:2012ufa} 
and so far, no signature of supersymmetry (SUSY) has been observed \cite{susyexp,Craig:2013cxa,Altmannshofer:2012ks}. Both of these, especially the second one, can have important repercussions on the implementation of HSMU.

In the earlier works on HSMU hypothesis, the issue of variation of SUSY breaking scale as a function of $\tan \beta$ was explored in the split SUSY scenario \cite{Mohapatra:2005pw}. In the present work, we derive new 
constraints on the allowed ranges of SUSY breaking scale and $\tan \beta$ in the case of Minimal Supersymmetric Standard Model (MSSM). It was also shown that this hypothesis works for a wide range of unification scales \cite{Mohapatra:2003tw}. We investigate 
it further and derive  new constraints on the variation of unification scale.  In view of the availability of more precise values of the neutrino oscillation parameters \cite{Fogli:2012ua, GonzalezGarcia:2012sz} these 
investigations are likely to serve as important tests of HSMU hypothesis. A detailed discussion of these constraints is one of the main features of this paper.

This paper is organized in the following manner. In section \ref{sec2}, we provide a general formalism of the RG running of Majorana neutrino masses and mixing angles. Section \ref{sec3}, contains our results for 
the neutrino oscillation parameters at low energy within the framework of HSMU hypothesis. In section \ref{sec4}, we discuss various predictions originating from our analysis.  The constraints on the unification scale, 
SUSY breaking scale and $\tan \beta $ are derived in section \ref{sec5}. Finally,  in section \ref{sec6}, we summarize our results and give future directions.


\section{Renormalization Group Evolution of Neutrino Masses and mixing angles}
\label{sec2}


We present, in this section, the RG equations used in our analysis. Our basic assumption is that the neutrinos are Majorana type and mass eigenstates $m_i$, $(i = 1, 2, 3)$ are of same CP parity. 
We also ignore CP violating phases in the mixing matrix. With these assumptions, the real PMNS matrix can be parametrized as 
\\
\noindent
\begin{equation}
U=\left[\br{ccc}
c_{12}c_{13}                       & s_{12}c_{13}                       & s_{13} \\
-s_{12}c_{23} - c_{12}s_{23}s_{13} & c_{12}c_{23} - s_{12}s_{23}s_{13}  & s_{23}c_{13} \\
s_{12}s_{23} - c_{12}c_{23}s_{13}  & -c_{12}s_{23} - s_{12}c_{23}s_{13} & c_{23}c_{13}
\er\right] ,
\end{equation} \\
with $c_{ij}=\cos\theta_{ij}$ and $s_{ij}=\sin\theta_{ij}$ $ (i,j = 1, 2, 3$). The  $U$ matrix diagonalizes the neutrino mass matrix $M$ in the flavor basis, i.e.  $U^TMU={\rm diag}(m_1, m_2, m_3)$.

Here, we take a model independent approach and assume that the new physics operating at the unification scale results in the unification between CKM and PMNS mixing angles. 
In order to get the low scale values, we work in type-I seesaw scenario. For the RG running from unification scale to seesaw scale, we use the standard MSSM RG equations 
within the framework of type-I seesaw mechanism \cite{Antusch:2005gp}. Below the seesaw scale all right handed neutrinos are integrated out and the masses of left handed 
neutrinos are generated by a dimension 5 operator added to the standard SM/MSSM Lagrangian. We have numerically checked our results by varying seesaw scale from $10^{13} - 10^9$ GeV 
and we find that our analysis depends weakly on the chosen value of seesaw scale.
Thus,  for the sake of illustration and definiteness, we have taken seesaw scale of ${\cal O} (10^{10})$ GeV throughout this work. 

At this juncture, we would like to point out that, for our analysis, we do not need any details of the theory operating at the unification scale. 
Although one such high energy theory has already been discussed in literature (see, e.g. \cite{Mohapatra:2003tw} for details). Moreover, RG equations presented here are at one loop 
level and only dominant terms are shown (cf. (2) and (5) below). However, in numerical computations, we have used full two-loop RG  equations \cite{Antusch:2002ek}.

The RG evolution of neutrino masses $m_i$, below seesaw scale, is determined by the following equations \cite{Casas:1999tg,Antusch:2002ek,Antusch:2003kp,Antusch:2005gp}
\begin{eqnarray}\label{rgmass}
 \frac{\D m_i}{\D t}
 & = & \frac{m_i }{16\pi^2}
        \left[
        \alpha + C f_\tau^2 \; F_i 
        \right] \;,
\end{eqnarray}
where $t=\ln(\mu/\mu_0)$, $\mu$ is the renormalization scale and \(F_i\) (with $i = 1, 2, 3$) are defined as
\begin{eqnarray}
        F_1 &=& 2 s_{12}^2 \, s_{23}^2  -s_{13} \, \sin 2\theta_{12} \, \sin 2\theta_{23}  +
        2 s_{13}^2 \, c_{12}^2 \, c_{23}^2 \;,  \nonumber \\
        F_2 &=& 2 c_{12}^2 \, s_{23}^2 + 
    s_{13} \, \sin 2\theta_{12} \, \sin 2\theta_{23}  +
        2 s_{13}^2 \, s_{12}^2 \, c_{23}^2 \;,  \nonumber\\
F_3 &=& 2 \, c_{13}^2 \, c_{23}^2.
\end{eqnarray}

In  SM and MSSM, \(\alpha\), $f_\tau$ and C are
\begin{eqnarray}
 \alpha_\mathrm{MSSM}
& = &
 -\frac{6}{5} g_1^2 - 6 g_2^2 + \frac{6 y_t^2}{\sin^2\beta} \;, \nonumber \\
         \alpha_\mathrm{SM}
& = &
        -3 g_2^2 + 2 y_\tau^2 
 +
        6 \left( y_t^2 + y_b^2  \right) 
        + \lambda \;, \nonumber \\
f_{\tau,\mathrm{MSSM}}^{2} & = & \frac{y^2_\tau}{\cos^2\beta}    \;, \qquad 
f_{\tau,\mathrm{SM}}^{2}  =  y^2_\tau \;, \nonumber\\
C &=& 1 \hphantom{-\frac{3}{2}} \;\; \text{in MSSM}\;, \qquad  C = -\frac{3}{2} \hphantom{1} \;\; \text{in SM}\;.
\end{eqnarray}
Here $ y_f$, $(f = \tau, t, b)$ represents the Yukawa coupling for $\tau$-lepton, top and bottom quarks, respectively.  
The gauge couplings are denoted by \(g_i\) and \(\lambda\) stands for the Higgs self-coupling in SM.  

The RG equations which govern evolution of mixing angles are given as \cite{Casas:1999tg,Antusch:2002ek,Antusch:2003kp,Antusch:2005gp}
\begin{eqnarray}
\label{thetarg}
\frac{\D \theta_{12}}{\D t}& = & -\frac{C f_\tau^2}{32\pi^2} \, \sin 2\theta_{12} \, s_{23}^2 \, \frac{ ({m_1} + {m_2})^2}{ \Delta m^2_{21}}   + \mathscr{O}(\theta_{13}),  \nonumber \\
\frac{\D \theta_{13}}{\D t} & = &  -\frac{C f_\tau^2}{32\pi^2} \,  \sin 2\theta_{12} \, \sin 2\theta_{23} \, \frac{m_3 }{\Delta m^2_\mathrm{32} \left( 1+\xi \right)}
        \left[ (m_2  - m_1 ) + \xi \left( m_2  + m_3 \right)  \right]    + \mathscr{O}(\theta_{13}),  \nonumber \\
\frac{\D \theta_{23}}{\D t} & = &  -\frac{C f_\tau^2}{32\pi^2} \, \sin 2\theta_{23} \, \frac{1}{\Delta m^2_\mathrm{32}} \left[ c_{12}^2 \, (m_2 + m_3)^2 + s_{12}^2 \, \frac{(m_1 + m_3)^2}{1+\xi}
        \right] 
          + \mathscr{O}(\theta_{13}),
\end{eqnarray}

with
\begin{eqnarray}
\xi &=&\,\frac{\Delta m^2_\mathrm{21}}{\Delta m^2_\mathrm{32}},  \qquad  \Delta m^2_\mathrm{21} = m_2^2-m_1^2, \qquad \Delta m^2_\mathrm{32} = m_3^2-m_2^2.
\end{eqnarray}

In this work, Dirac as well as  Majorana phases of the PMNS mixing matrix are taken to be zero.  The results with non-zero phases will be presented in a future publication \cite{AGRS}.  
In (\ref{rgmass}) and (\ref{thetarg}), for sake of brevity, we have given only the dominant terms of the RG equations at one loop level. 
The full two loop RG equations, used in this work, can be found in \cite{Antusch:2002ek}. The numerical computations, at two loop, are done using a MATHEMATICA based package REAP \cite{Antusch:2005gp}.


\section{Magnification of Mixing Angles via RG Evolution }
\label{sec3}


The HSMU hypothesis is implemented in two steps.  We first follow a bottom-up approach and take the known values of gauge couplings, Yukawa couplings and
CKM matrix elements at the low scale $(M_Z)$ \cite{Xing:2011aa} and  evolve them up to  the SUSY breaking scale $(M_{SUSY})$ using the standard SM RG equations \cite{Antusch:2003kp}.  
From the SUSY breaking scale to the unification scale, the evolution of these parameters is governed by MSSM RG equations \cite{Antusch:2002ek,Antusch:2003kp}. 

At the unification scale, following the HSMU hypothesis, we assume that the PMNS mixing angles ($\theta_{12}^0,\theta_{13}^0, \theta_{23}^0 $) are identical to the CKM mixing angles ($\theta_{12}^{0,q},\theta_{13}^{0,q}, \theta_{23}^{0,q}$). 
In addition to this, we choose initial neutrino masses to be quasi-degenerate with normal hierarchy pattern and PMNS phases to be zero.
The requirements of normal hierarchy and quasi-degeneracy of neutrinos are essential ingredients to achieve large mixing angle magnification (within the 3-$\sigma$ range at the low scale) \cite{Mohapatra:2003tw}. 
 
We then follow a top-down approach and run down the neutrino masses and mixing angles from unification scale to the seesaw scale using MSSM RG equations within the framework of type-I seesaw mechanism \cite{Antusch:2005gp}.
From seesaw scale to SUSY breaking scale, the running is done using MSSM RG equations with dimension-5 operator \cite{Antusch:2002ek,Antusch:2003kp}.  
Below the SUSY breaking scale to the low scale, RG running is governed by the  SM RG equations.

In the earlier works on  HSMU hypothesis \cite{Mohapatra:2003tw,Mohapatra:2004vw,Mohapatra:2005gs,Mohapatra:2005pw,Agarwalla:2006dj}, the SUSY breaking scale  was taken as 1 $\rm TeV$.  
At present, this is not favored by direct SUSY searches at the LHC \cite{susyexp,Craig:2013cxa}. In view of this,  we have taken  the SUSY breaking scale as $2$ TeV. 
The working of HSMU hypothesis requires large values of $\tan \beta$ which is also consistent with constraints imposed by SUSY searches 
\cite{Craig:2013cxa,Altmannshofer:2012ks,Beskidt:2013gia,Abada:2012re,Hirsch:2012ti}. 
Therefore, in this section, we have taken $\tan \beta$ to be 55. Moreover, we have taken unification scale to be $2 \times 10^{16}$ GeV which is a generic scale for Grand Unified Theories  (GUTs). 
The dependence of our analysis on these parameters is discussed in section \ref{sec5}. 

\begin{table*} [h!]
\caption{Radiative magnification to bilarge mixings at low energies for input values of $\theta_{12}^0 = \theta_{12}^{0,q}= 13.02^0$, $\theta_{23}^0 = \theta_{23}^{0,q}
=2.03^0$ and $\theta_{13}^0=\theta_{13}^{0,q}= 0.17^0$.  We have taken the unification scale $= 2 \times 10^{16}$ GeV, $M_{SUSY}$ = 2 TeV and tan $\beta = 55$. The various 
entries in the table also highlight the correlations between low scale neutrino oscillation parameters.}
\begin{ruledtabular}
\begin{tabular}{lcccccc}
                                                &           I                 &             II                &             III              &              IV               &           V                      &\\ \hline
$m_1^0$(eV)                                     & 0.4152                      &  0.3972                       &  0.4344                      & 0.4102                        &  0.4240                          &\\
$m_2^0$(eV)                                     & 0.4186                      &  0.4005                       &  0.4380                      & 0.4137                        &  0.4275                          &\\
$m_3^0$(eV)                                     & 0.4825                      &  0.4617                       &  0.5049                      & 0.4769                        &  0.4928                          &\\
$m_1$(eV)                                       & 0.3577                      &  0.3422                       &  0.3742                      & 0.3534                        &  0.3653                          &\\
$m_2$(eV)                                       & 0.3583                      &  0.3428                       &  0.3749                      & 0.3541                        &  0.3659                          &\\
$m_3$(eV)                                       & 0.3620                      &  0.3463                       &  0.3788                      & 0.3578                        &  0.3697                          &\\
$\Dmott$(eV$^2$)$_{RG}$                         & $4.29 \times 10^{-4}$       &  $3.93 \times 10^{-4}$        & $4.70\times 10^{-4}$         & $4.49\times 10^{-4}$          &  $4.20\times 10^{-4}$            & \\
$\Dmttht$(eV$^2$)$_{RG}$                        & $2.67 \times 10^{-3}$       &  $2.45\times 10^{-3}$         & $2.92 \times 10^{-3}$        & $2.61 \times 10^{-3}$         &  $2.78 \times 10^{-3}$           & \\
$M_{\tilde e}/M_{\tilde \mu, \tilde \tau}$      & 1.85                        & 1.81                          & 1.89                         & 1.76                          &  2.06                            &\\
$\Dmott$(eV$^2$)$_{th}$                         & $-3.54\times 10^{-4}$       & $-3.12\times 10^{-4}$         & $-4.00 \times 10^{-4}$       & $-3.73 \times 10^{-4}$        &  $-3.44 \times 10^{-4}$          & \\
$\Dmttht$(eV$^2$)$_{th}$                        & $-2.74\times 10^{-4}$       & $-2.41\times 10^{-4}$         & $-3.09 \times 10^{-4}$       & $-2.16\times 10^{-4}$         &  $-3.81\times 10^{-4}$           & \\
$\Dmott$(eV$^2$)                                & $7.52 \times 10^{-5}$       & $8.07 \times 10^{-5}$         & $7.02 \times 10^{-5}$        & $7.57 \times 10^{-5}$         &  $7.56 \times 10^{-5}$           & \\
$\Dmttht$(eV$^2$)                               & $2.40 \times 10^{-3}$       & $2.20\times 10^{-3}$          & $2.62\times 10^{-3}$         & $2.40\times 10^{-3}$          &  $2.40\times 10^{-3}$            & \\
$\theta_{23} /^\circ$                           & $54.00$                     & $54.00$                       & $54.00$                      & $53.84$                       &  $54.10$                         & \\
$\theta_{13} /^\circ$                           & $8.67$                      & $8.67$                        & $8.67$                       & $8.66$                        &  $8.66$                          & \\
$\theta_{12} /^\circ$                           & $33.38$                     & $33.38$                       & $33.38$                      & $31.14$                       &  $35.87$                         & \\
\end{tabular} 
\end{ruledtabular}
  \label{tab2}
\end{table*}

In Table \ref{tab2}, we present five sets of neutrino oscillation parameters at low and high energy scales obtained within HSMU hypothesis. Each column in the table  depicts some specific set of values for neutrino 
oscillation parameters chosen in a way to show correlations between them. In order to highlight the correlation between any two low scale parameters we choose the unification 
scale neutrino masses such that all other parameters, at the low scale, remain close to their best fit values\footnote{The RG evolution of $\theta_{13}$ and $\theta_{23}$ is correlated in the HSMU hypothesis. 
Therefore, at the low scale, both cannot be obtained near their best fit values simultaneously.}.  In column I, all the low scale parameters are obtained close to their best fit values, 
except $\theta_{23}$ which is $54^\circ$.  In column II, keeping $\theta_{13}$ and $\theta_{12}$ close to their best fit values at the low scale, the values of $\Dmott$ and $\Dmttht$ are 
obtained at their 3-$\sigma$ upper and lower edge respectively. For this pattern, $\theta_{23}$ again turns out to be $54^\circ$, i.e. non-maximal. Whereas, in column III,  $\Dmott$ and 
$\Dmttht$  are respectively kept at their 3-$\sigma$ lower and upper edge.  The rest of the results are similar to the previous ones. In columns IV and V,  $\theta_{12}$ is taken to its 
lower and upper 3-$\sigma$ limit, respectively, keeping all other parameters (except $\theta_{23}$) close to their best fit values at the low scale. We see that $\theta_{23}$ always remains 
above $45^\circ$ and lies in the second octant. Moreover, as is clear from Table \ref{tab2}, for a fixed value of $\theta_{13}$, the correlation between $\theta_{12}$ and $\theta_{23}$ is weak.

\begin{figure}[h!]
 	\begin{center}\vspace{1.25cm}
 	 \includegraphics[width = 12.0cm]{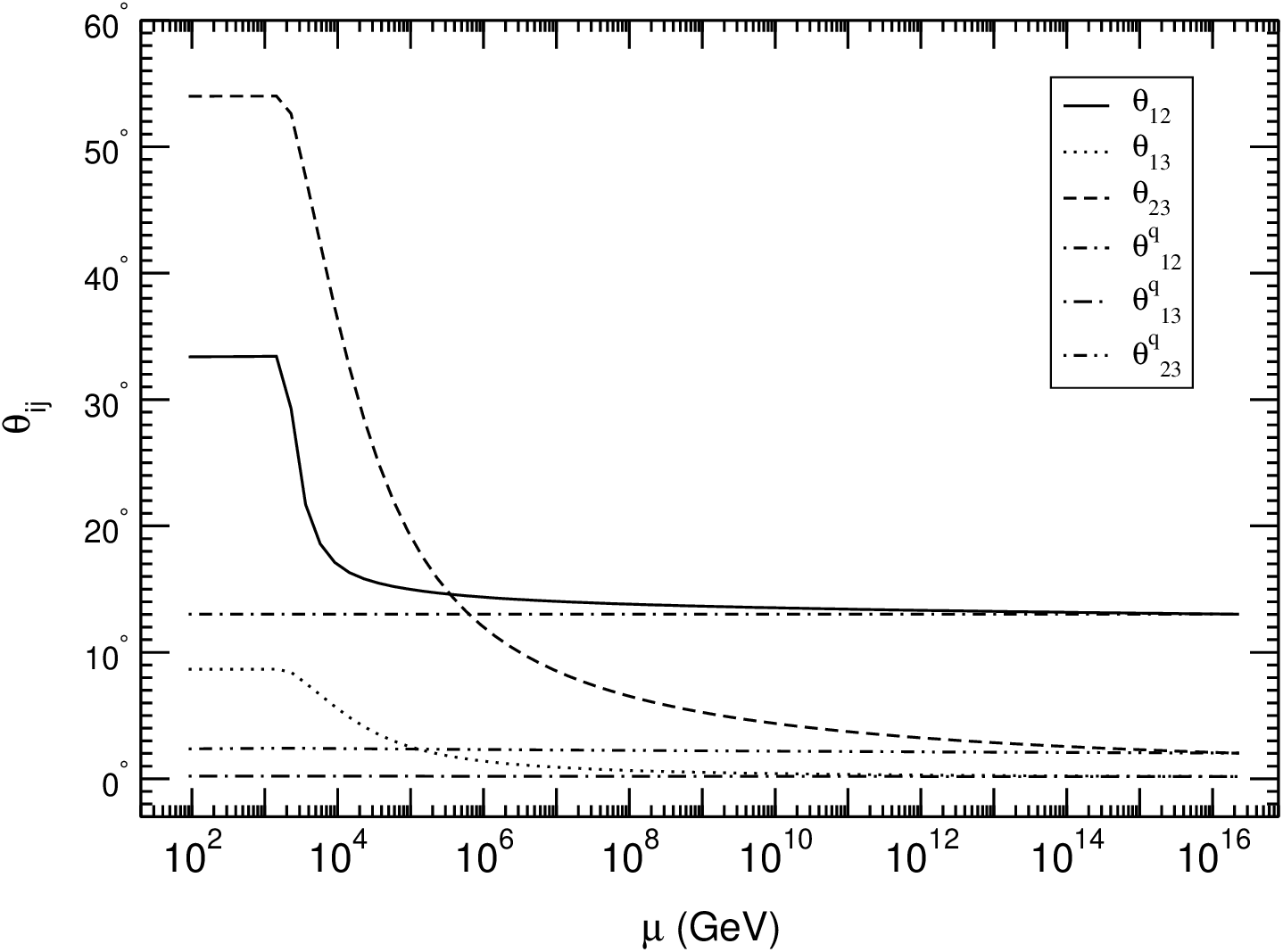}
	\caption{ The RG evolution of CKM and PMNS mixing angles with respect to RG scale $(\mu)$. This figure corresponds to the neutrino oscillation parameters quoted in the first column of Table \ref{tab2}.}
	\label{fig1}
 	\end{center}\vspace{0.125cm}
\end{figure}

The RG evolution of the three PMNS and CKM mixing angles from the unification scale ($2 \times 10^{16}$ GeV) to the low scale $(M_Z)$ is shown in Figure \ref{fig1}. 
As clear from the figure, owing to the quasi-degeneracy of neutrino masses, large angle magnification occurs in the PMNS sector. The magnification of CKM mixing angles $(\theta^q_{ij}, \; i, j = 1,2,3)$ is
almost negligible because of the hierarchical nature of quark masses. We also observe that the major part of magnification occurs near  SUSY breaking scale which, 
in this case, is chosen to be $M_{SUSY} = 2 \times 10^3$ GeV.  The SM RG equations lead to negligible angle magnification as clear from the flatness of curves below $M_{SUSY}$.

\begin{figure}[h!]
 	\begin{center}\vspace{1.2cm}
 	 \includegraphics[width = 12.0 cm]{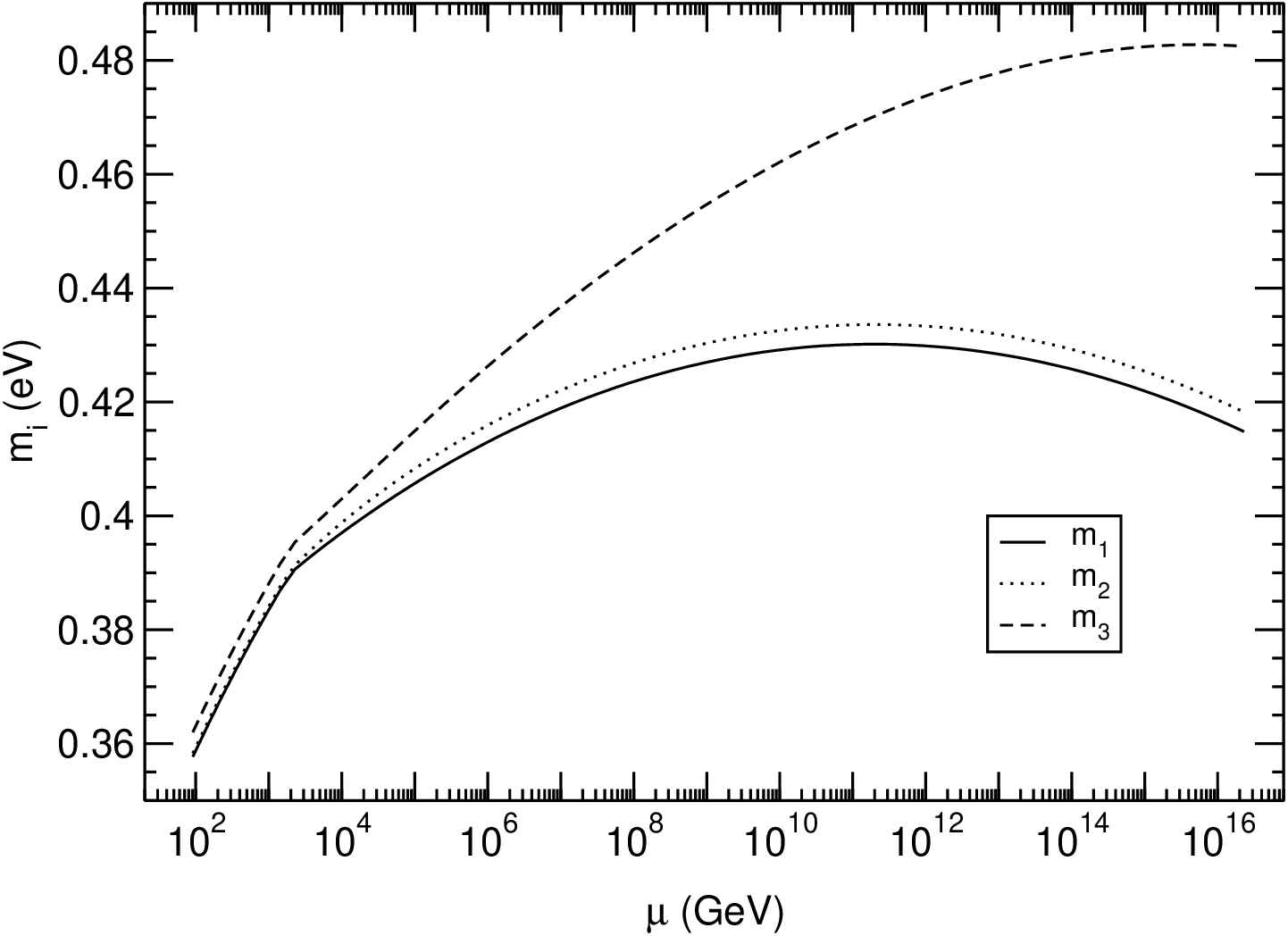}
	\caption{ The RG evolution of neutrino masses $(m_i)$ with respect to RG scale $(\mu)$. This figure corresponds to the values in the first column of Table \ref{tab2}.  }
	\label{fig2}
 	\end{center}\vspace{0.125cm}
\end{figure}

The RG evolution of neutrino masses from unification scale to $M_Z$ is shown in Figure \ref{fig2}. It is clear that all the masses decrease as we move from unification scale to low scale (cf. Figure \ref{fig2}). 
Initially, at unification scale, the splitting among the masses is relatively large but after RG evolution the splitting gets narrowed down and they acquire nearly degenerate mass at $M_Z$.

\subsection*{Low energy threshold corrections to neutrino masses}


It is evident from table \ref{tab2} that only one (i.e. $\Delta m^2_{32}$) out of two mass squared differences, at the low scale, lies within experimental 3-$\sigma$ range. This discrepancy can easily be accounted for by
threshold corrections \cite{Mohapatra:2005gs,Agarwalla:2006dj}.  In the case of quasi-degenerate  neutrinos, the low energy MSSM threshold corrections can result in a significant contribution, as shown in 
\cite{Chun:1999vb,Chankowski:2000fp,Chun:2001kh,Chankowski:2001hx}. These threshold corrections are given by following equations \cite{Mohapatra:2005gs,Agarwalla:2006dj}: 
\begin{eqnarray} 
(\Delta m_{21}^2)_{th}
&=& 2\rm m^2\cos 2\theta_{12}[-2T_e + T_\mu + T_\tau], \nonumber \\
(\Delta m_{32}^2)_{th}
&=& 2\rm m^2\sin^2\theta_{12}[-2T_e + T_\mu + T_\tau], \nonumber \\
(\Delta m_{31}^2)_{th}
&=& 2\rm m^2\cos^2\theta_{12}[-2T_e + T_\mu + T_\tau].
\end{eqnarray}
Here, $m = \frac{1}{3}(m_1 + m_2 + m_3)$ is the mean mass of the quasi-degenerate neutrinos and $T_{\alpha} (\alpha = e, \mu, \tau)$ is the one-loop factor. 
Its form has been previously calculated in  \cite{Chankowski:2001hx,Chun:1999vb} and given by
\begin{eqnarray}
T_{\alpha}&=& \frac{g^2_2}{32\pi^2} \left[\frac{x_{\mu}^2-x_{\alpha}^2}{y_{\mu}y_{\alpha}} + \frac{(y_{\alpha}^2-1)}{ y_{\alpha}^2}ln(x_{\alpha}^2) 
- \frac{(y_{\mu}^2-1)}{y_{\mu}^2} ln(x_{\mu}^2) \right],
\label{eq23}
\end{eqnarray}
\par
\noindent
where $g_2$ is the $SU(2)$ coupling constant and $y_{\alpha} = 1-x_{\alpha}^2$ with $x_{\alpha} = M_{\alpha}/M_{\tilde w}$; $M_{\tilde w}$ stands for wino mass, $M_{\alpha}$ represents the mass of charged sleptons.
Moreover, without any loss of generality, the loop-factor has been defined to give $T_{\mu} = 0$ (cf. \cite{Mohapatra:2005gs,Agarwalla:2006dj} for details). 

At the LHC, for simplified scenarios, chargino masses are excluded up to $750$ GeV in the presence of light sleptons and up to $300$ GeV in the case of heavy sleptons  \cite{susyexp,Craig:2013cxa}. 
In the view of above constraints, here we have taken the wino mass to be $800$ GeV.

After the inclusion of threshold corrections, along with the RG-evolution effects, the final expression for mass squared differences is given as 

\noindent
\begin{equation}
 \Delta m_{ij}^2=(\Delta m_{ij}^2)_{\rm RG} +  (\Delta m_{ij}^2)_{\rm th} \label{eq25}.
\end{equation}

It is clear from table \ref{tab2}, that the RG effects, along with threshold corrections, result in good agreement between the predictions of HSMU hypothesis and the 
present experimentally allowed range of neutrino oscillation parameters (cf. Table \ref{tab1}). At this juncture, we would like to point out that, although the threshold corrections  
for mass square  differences are significant yet they are negligibly  small compare to the mean mass of neutrinos. The same is also true for the threshold corrections to mixing angles \cite{Mohapatra:2005gs,Agarwalla:2006dj}.


\section{Predictions from HSMU hypothesis}
\label{sec4}

Within the framework of the HSMU hypothesis, the low energy oscillation data can be used to put stringent constraints on the allowed parameter range for the neutrino masses and mixing angle. 
The aim of this section is to discuss the predictions from our analysis which are obtained after imposing these constraints. These predictions can be tested in present and future experiments
as discussed in this section.


\subsection{Predictions for Masses,  $\langle M_{\beta} \rangle$ and  $\langle M_{\beta\beta} \rangle$ at $M_Z$}
\label{subsec32}

As clear from Table \ref{tab2}, the neutrino masses at $M_Z$ lie in the range of $0.34$-$0.38$ eV.  This range can be probed by various presently running as well as near future experiments and 
hence it provides an important test for HSMU hypothesis.  For example, the recent result from GERDA gives an upper bound of $0.2$-$0.4$ eV on the 
$\langle M_{\beta\beta} \rangle$ component of mass matrix \cite{Agostini:2013mzu}. Similarly, EXO-200 provides an upper bound of $0.14$-$0.38$ eV  on the 
same \cite{Auger:2012ar}. Although the present bounds on $\langle M_{\beta} \rangle$ from tritium beta decay are comparatively weak ($<$ 2 eV) \cite{Kraus:2004zw,Aseev:2011dq,Beringer:1900zz}.
In future, the KATRIN experiment will be able to probe it down to 0.2 eV \cite{Drexlin:2013lha}. 

Moreover, the recent Planck data has provided a bound on the sum of neutrino masses in the range of  $0.23$-$1.08$ eV depending on the choice of 
the priors  \cite{Ade:2013zuv}. The lower limit of Planck is in tension with our hypothesis  but it should be noted that the cosmological constraints are highly model dependent and should 
be taken in conjunction with other experiments. In view of the above considerations, the absolute value of neutrino masses provides an important test of our hypothesis.  We would like 
to point out that the above mentioned mass range ($0.34$-$0.38$ eV) is obtained for a specific choice of unification scale, SUSY breaking scale and $\tan \beta$ (cf. Table \ref{tab2} for details). 
The dependence of neutrino masses (at $M_Z$) with respect to these parameters is discussed, in detail, in section \ref{sec5}.


\subsection{Predictions for mixing angles at $M_Z$}
\label{subsec33}

\begin{figure}[h!]
 	\begin{center}\vspace{1.25cm}
 	 \includegraphics[width = 12.0 cm]{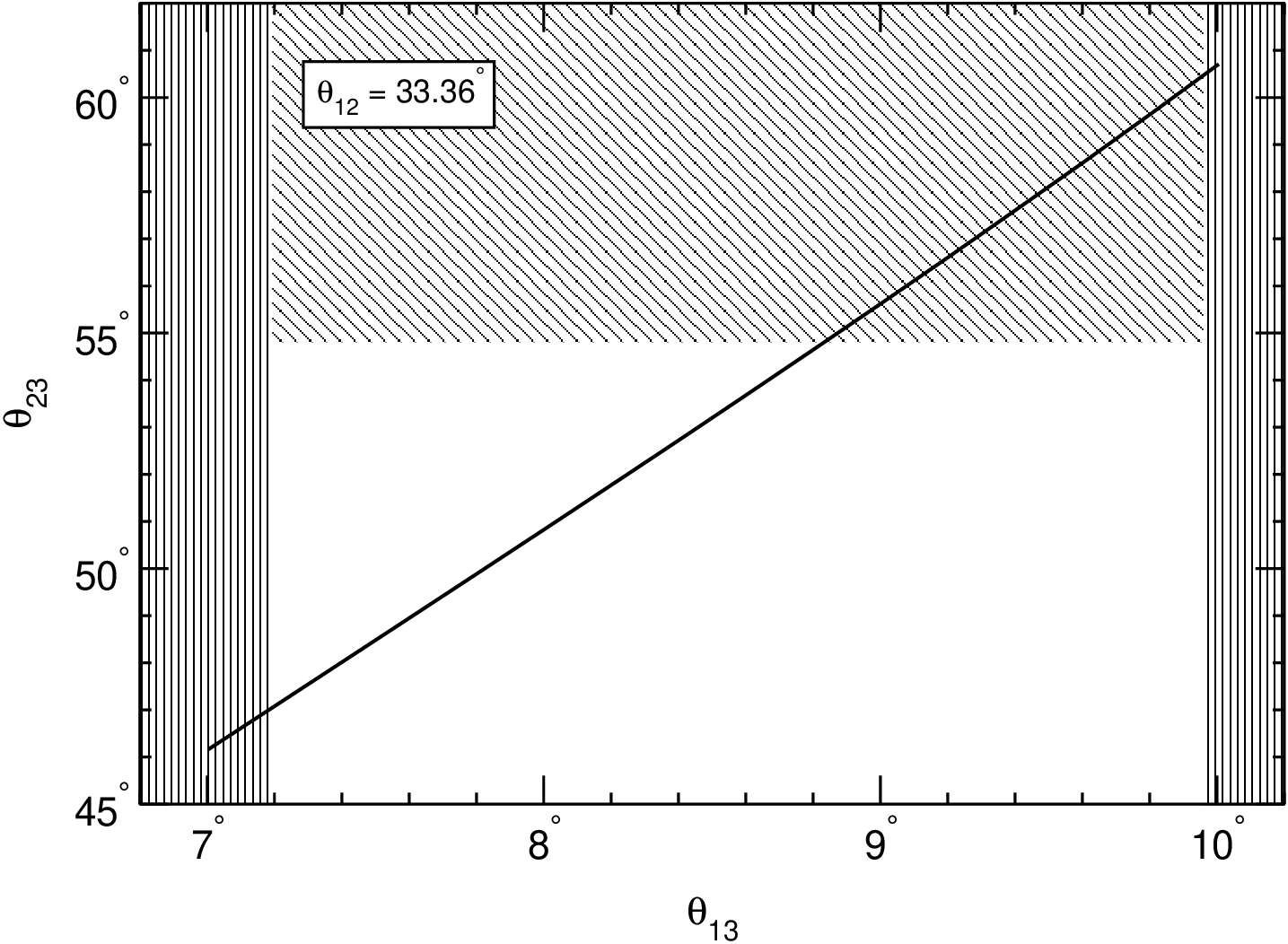}
	\caption{ The variation of $\theta_{23}$ with respect to $\theta_{13}$. For plotting this figure we have kept all other oscillation parameters to be at their best-fit values. 
         The vertically shaded regions lie outside the 3-$\sigma$ range of $\theta_{13}$ whereas the horizontally shaded 
         one lies outside 3-$\sigma$ range of $\theta_{23}$ \cite{GonzalezGarcia:2012sz}. }
	\label{fig3}
 	\end{center}\vspace{0.1cm}
\end{figure}

 It is clear from the RG equations (\ref{thetarg}) that, within HSMU hypothesis, the mixing angles $\theta_{13}$ and $\theta_{23}$ are correlated.   
In Figure \ref{fig3}, we show the explicit dependence of $\theta_{23}$  on  $\theta_{13}$ 
keeping other low scale neutrino oscillation parameters fixed near to their best fit values.  We  observe that $\theta_{23}$ turns out to be above $45^\circ$ (i.e. lies in the second octant), 
for the whole 3-$\sigma$ range of $\theta_{13}$. This prediction is easily  testable in the current and in future experimets, like INO, T2K, NO$\nu$A, LBNE, Hyper-K and PINGU 
\cite{Abe:2011ks,Patterson:2012zs,Adams:2013qkq,Ge:2013ffa,Kearns:2013lea,Athar:2006yb}.

Even for the lower edge value of the present 3-$\sigma$ range of $\theta_{13}$, the value of $\theta_{23}$ is non-maximal and is around $47^\circ$, as evident from Figure \ref{fig3}. The values of $\theta_{23}$ 
increase with $\theta_{13}$. When $\theta_{13}$ is around $9^\circ$, $\theta_{23}$ reaches its upper edge of 3-$\sigma$ limit and  it goes into the disfavored region for higher values of $\theta_{13}$ 
(which is still within its 3-$\sigma$ range).  This, in turn, puts constraints on the values of  $\theta_{13}$, which should lie in the range $7.19^\circ$-$8.8^\circ$\footnote{As shown in table \ref{tab2}, $\theta_{23}$ 
also depends very weakly on $\theta_{12}$.  The above quoted range is for $\theta_{12}$ at its best fit value. }.

At this point we would like to mention that, the RG evolution of $\theta_{12}$ also depends on $\Delta m^2_{21}$. 
Therefore, it can be varied independently of the other two angles by making an appropriate choice of $\Delta m^2_{21}$ at unification scale. Hence, within HSMU hypothesis, no effective constraints on its range can be obtained.


\section{Allowed Parameter Range for Unification Scale, SUSY Breaking Scale and $\tan\beta $}
\label{sec5}


In this section, we study the variation of unification scale, SUSY breaking scale and $\tan \beta$ and its impact on HSMU hypothesis. We derive constraints 
on the range of these parameters for which HSMU hypothesis works. For this purpose, in this section, we have fixed the values of experimentally measured quantities 
$\theta_{12}$ and $\theta_{13}$ to their best fit values (i.e. $33.36^\circ$ and $8.66^\circ$ respectively) at $M_Z$. We also fix $\Delta m^2_{\rm{32}} = 2.5 \times 10^{-3} \rm{eV}^2$,
which is slightly higher than its best fit value,  so that after adding appropriate threshold corrections, it remains within 3-$\sigma$ range. 

Since in our hypothesis  the quantities  $\theta_{23}$ and $\Delta m^2_{\rm{21}}$ are fixed in terms of other quantities, we have not put any restrictions on them, apart 
from the fact that, after adding appropriate threshold corrections they should remain within 3-$\sigma$ limit.


\subsection{Variation of Unification Scale}
\label{subsec41}


In the previous sections, we have chosen our unification scale as $2 \times 10^{16}$ GeV which is the typical scale for GUTs. Since our hypothesis  does not depend on the details of the high scale theory, it is 
not necessary to take the unification scale to be same as that of GUT. Thus, in this subsection, we analyze the effect of variation of unification scale. 

\begin{figure}[h!]
 	\begin{center}\vspace{1.25 cm}
 	 \includegraphics[width = 12.0 cm]{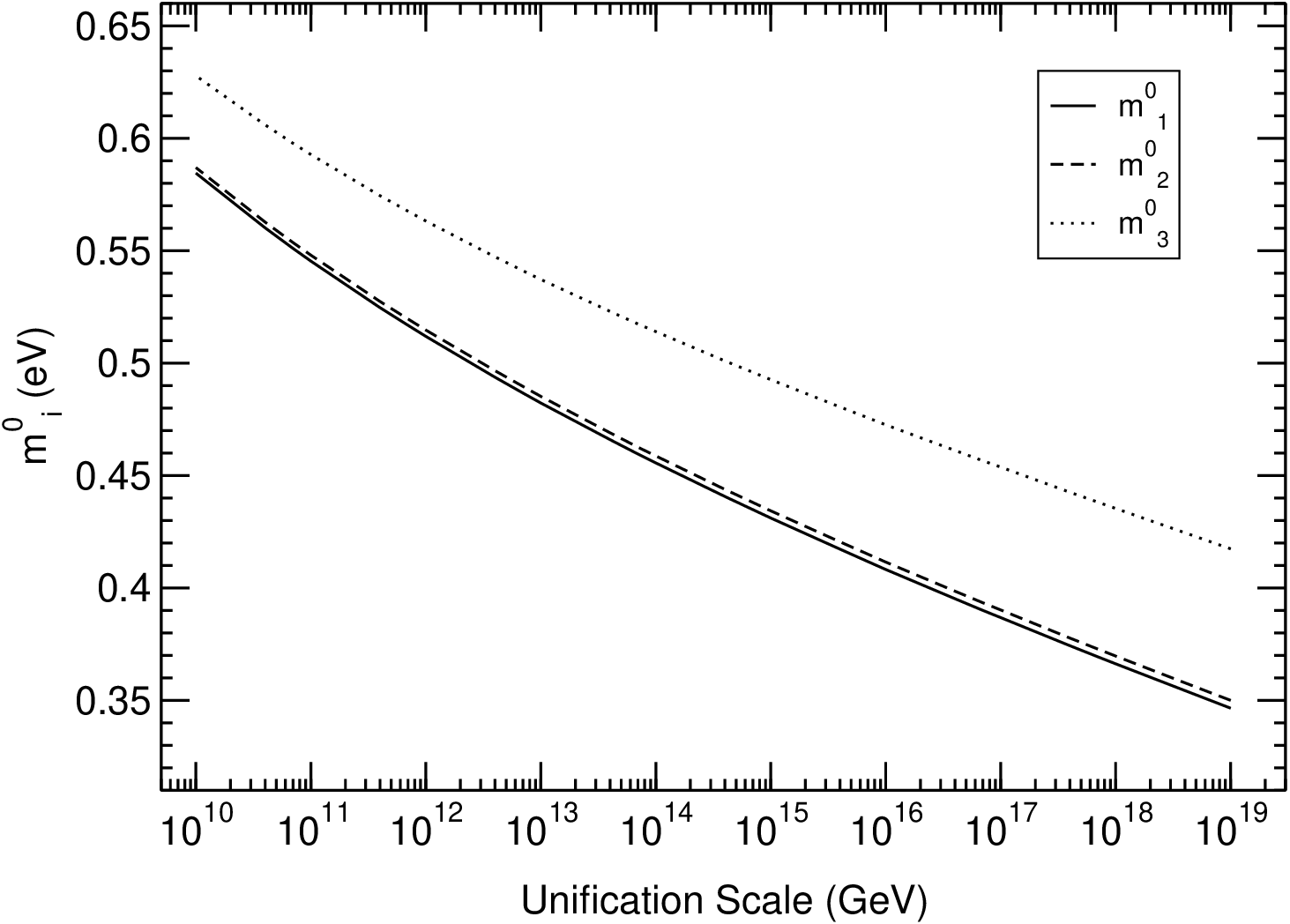}
	\caption{Unification scale vs neutrino masses $(m_i^0)$ at unification scale. In plotting this figure we have taken $M_{SUSY} = 2 \times 10^3$ GeV and $\tan \beta = 55$. }
	\label{fig4}
 	\end{center}\vspace{0.125cm}
\end{figure}

\begin{figure}[h!]
 	\begin{center}\vspace{1.25cm}
 	 \includegraphics[width = 12.0 cm]{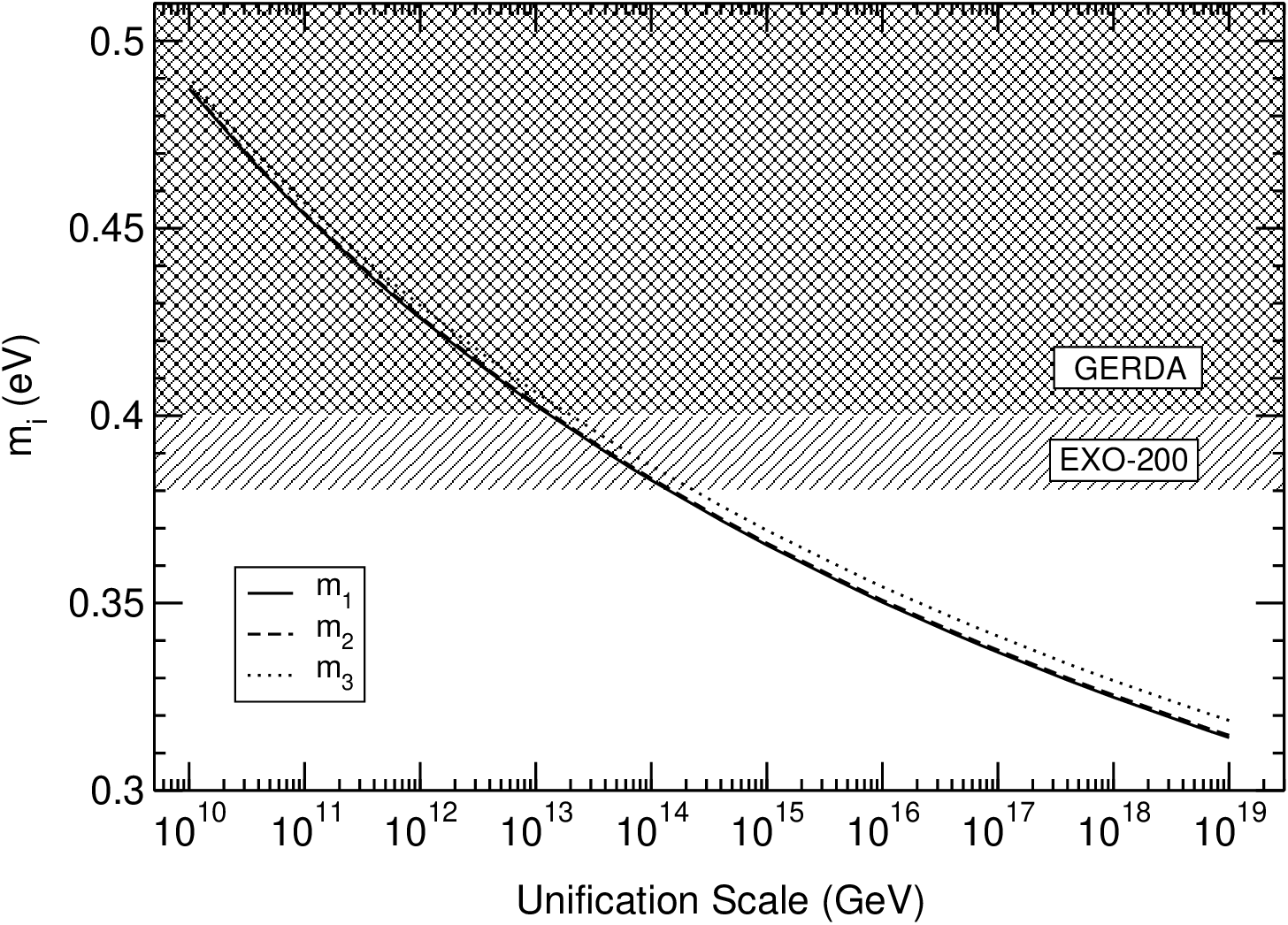}
	\caption{Unification scale vs neutrino masses $(m_i)$ at $M_Z$. Here we have taken $M_{SUSY} = 2 \times 10^3$ GeV and $\tan \beta = 55$. 
	The shaded regions are excluded by  $0\nu\beta\beta$ decay experiments \cite{Agostini:2013mzu,Auger:2012ar}.} 
	\label{fig5}
 	\end{center}\vspace{-0.125cm}
\end{figure}

It is clear from Figure \ref{fig1}, that a major part of angle magnification happens only close to $M_{SUSY}$. Therefore, it is expected that the desired angle magnification can be achieved 
even when the unification scale is not same as the GUT scale. In Figure \ref{fig4} and \ref{fig5}, we have, respectively, shown the 
variation of unification scale with respect to high and low scale neutrino masses. The magnitude of low scale masses (and derived quantities such as  $\langle M_{\beta} \rangle$ and  $\langle M_{\beta\beta} \rangle$) 
put constraints\footnote{In our case, since the neutrinos are quasi-degenerate and phases are absent, the mean mass ($m$) and  $\langle M_{\beta\beta} \rangle$ 
        are almost the same. Hence, in drawing the constraints in Figures \ref{fig5}, \ref{fig7} and \ref{fig9} we have neglected the small difference in the exact values of $m$ and $\langle M_{\beta\beta} \rangle$.  }
 on the unification scale as evident from the Figure \ref{fig5} and further elaborated in Section \ref{subsec44}.

Our analysis works for a wide range of unification scale from the Planck scale to much lower scales (cf. Figures \ref{fig4}, \ref{fig5}). The reason for this is that the major part of magnification of angles 
happens in a relatively small range near $M_{SUSY}$. Hence, one can take the unification scale to be several orders of magnitude lower than the GUT scale and still achieve desired magnification at $M_Z$.
The noteworthy point is that as we lower the unification scale the input neutrino masses have to be taken more degenerate because the range of MSSM RG running becomes shorter (cf. Figure \ref{fig4}). Thus, 
to achieve desired magnifications at $M_Z$, one has to make the input neutrino masses more degenerate to account for the lesser range of MSSM RG running. 

This increasing degeneracy of masses, in turn, results in $\Delta m^2_{\rm{32}}$ approaching 
its 3-$\sigma$ range much before $M_Z$. Therefore, to counter this and to keep $\Delta m^2_{\rm{32}}$ within its 3-$\sigma$ range at $M_Z$, one is also forced to increase the mean input mass at unification scale.
Furthermore, once the input mean mass is increased, it results in a relative increase in the mean mass at $M_Z$, partly because now it is higher to begin with
and partly because of the small range of MSSM RG running. 

Thus, the mean mass of neutrinos at unification scale as well as at $M_Z$ increases as we decrease the unification scale.
Hence, one can constrain the lowest possible unification scale using data from various experiments. 
We will further elaborate on such experimental constraints in Section \ref{subsec44}.


\subsection{Variation of SUSY Breaking Scale }
\label{subsec42}


 \begin{figure}[h!]
 	\begin{center}\vspace{0.7cm}
 	 \includegraphics[width = 12.0 cm]{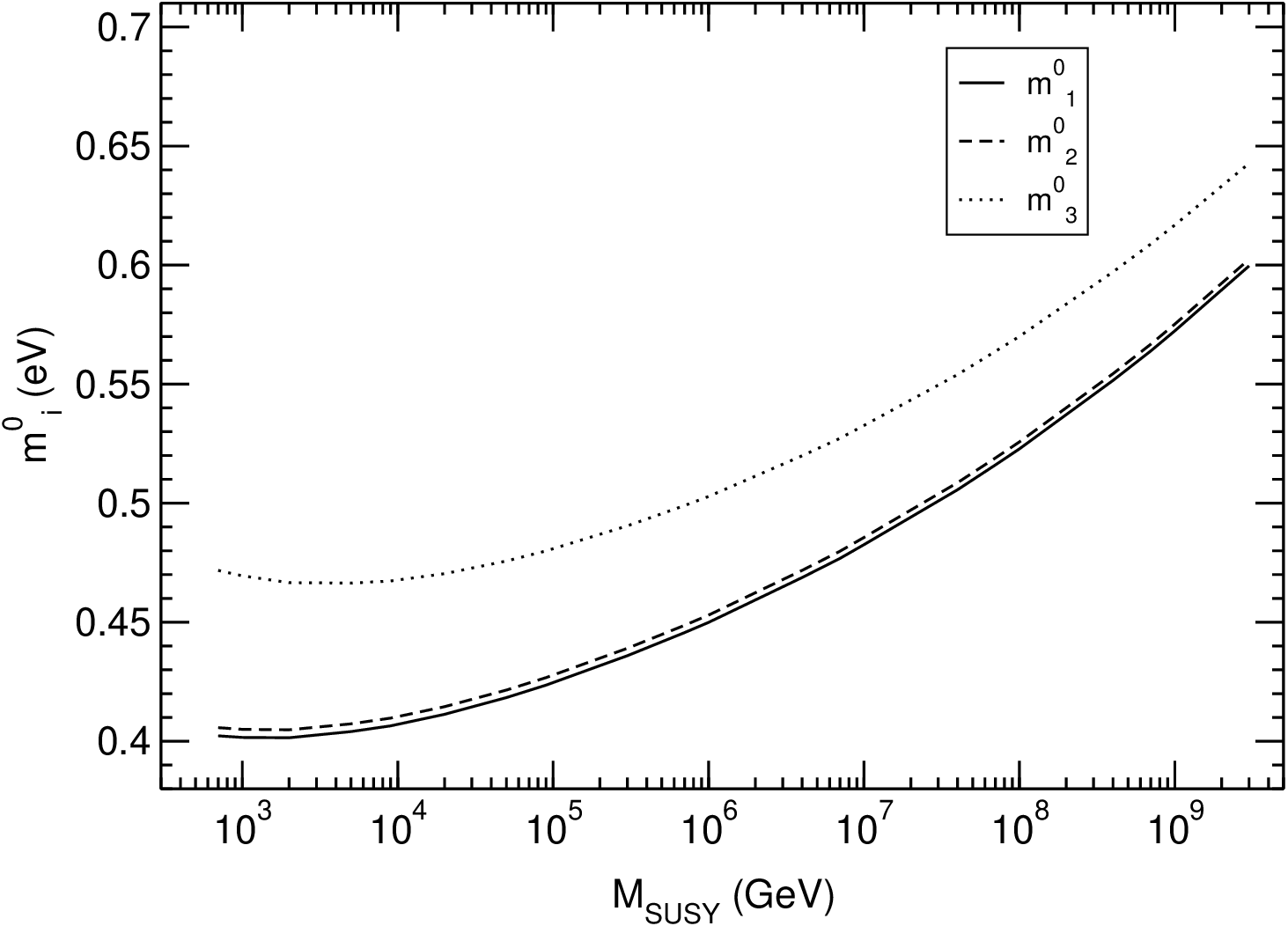}
	\caption{$M_{SUSY}$ vs neutrino masses $(m_i^0)$ at unification scale. In plotting 
                  this figure, we have taken unification scale $= 2 \times 10^{16}$ GeV and $\tan \beta = 55$. }
	\label{fig6}
 	\end{center}\vspace{0.cm}
\end{figure}

\begin{figure}[h!]
 	\begin{center}\vspace{1cm}
 	 \includegraphics[width = 12.0 cm]{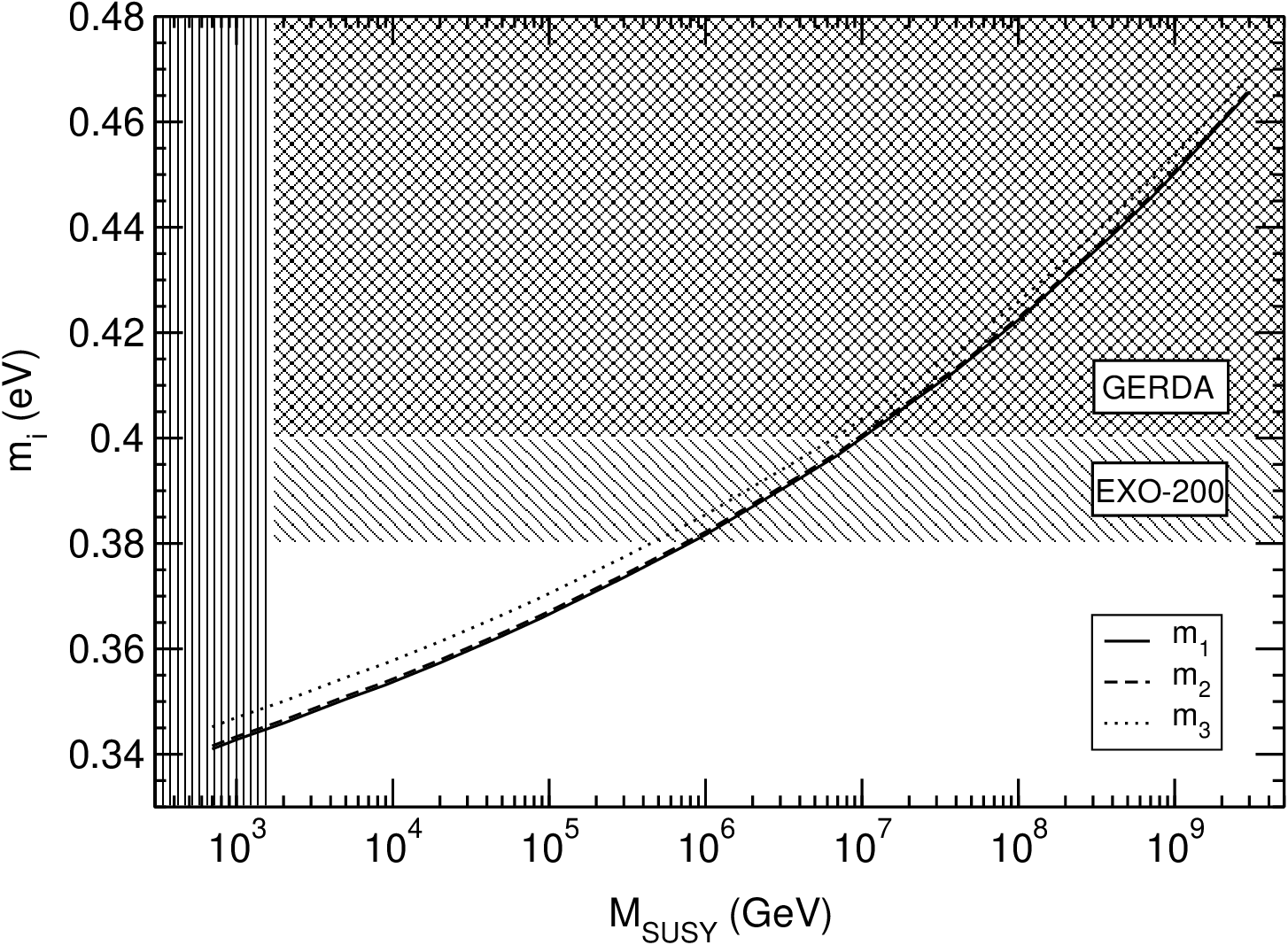}
	\caption{ $M_{SUSY}$ vs neutrino masses $(m_i)$ at $M_Z$. In plotting 
                  this figure, we have taken unification scale $= 2 \times 10^{16}$ GeV and $\tan \beta = 55$.  
                  The vertically shaded region is disfavored by the LHC SUSY searches \cite{Craig:2013cxa} whereas the 
                 horizontal ones are excluded by $0\nu\beta\beta$ decay experiments \cite{Agostini:2013mzu,Auger:2012ar}.  }
	\label{fig7}
 	\end{center}\vspace{0.125cm}
\end{figure}

We have, so far, fixed the SUSY breaking scale at $2 \times 10^3 $ GeV. In this section, we analyze the effects of variation of SUSY breaking scale. It is clear from Figure \ref{fig1} that the major 
part of magnification occurs only in and around the SUSY breaking scale. So one should expect to shift the scale of SUSY breaking from the so far chosen value and still be able to achieve the desired magnification.  
Our analysis works for a wide range of SUSY breaking scale starting from the TeV scale to much higher scales (as is clear from Figures \ref{fig6} and \ref{fig7}). 
While plotting these figures, we have taken the unification scale $= 2 \times 10^{16}$ GeV, $\tan \beta = 55$ and the value of 
observables at $M_Z$ to be same as before. 

As we increase the SUSY breaking scale, the input neutrino masses have to be taken to be more degenerate. The reason for this is that by increasing the SUSY breaking scale the range of MSSM RG running becomes shorter.
Thus, to achieve desired magnifications at $M_Z$ one has to make the input neutrino masses more degenerate in order to counter the lesser range of MSSM RG running. At the same time, we have to increase the mean mass of 
neutrinos at unification scale in order to keep the $\Delta m^2_{\rm{32}}$ within its 3-$\sigma$ range at $M_Z$. 

Since the mean mass of the neutrinos increases with increasing SUSY breaking scale, one can constrain the highest possible SUSY breaking scale using data from various experiments. 
Moreover, the lower ranges of SUSY breaking scale are constrained from SUSY searches at the LHC \cite{susyexp,Craig:2013cxa}. We further discuss these constraints in the Section \ref{subsec44}.


\subsection{Variation of $\tan\beta $}
\label{subsec43}


In MSSM, the RG running of angles gets enhanced by a factor of $(1 + \tan^2\beta)$ [cf. (\ref{thetarg}) for details]. Therefore, the larger values of $\tan\beta$ enhance the magnification at $M_Z$. This is the reason for choosing
$\tan\beta = 55 $ in the previous sections of this work.  We have, so far, fixed $\tan\beta = 55$ but in this section we will vary $\tan\beta $ to obtain the lower limits on it for desired magnification. \\

\begin{figure}[h!]
 	\begin{center}\vspace{0.5cm}
 	 \includegraphics[width = 12.0 cm]{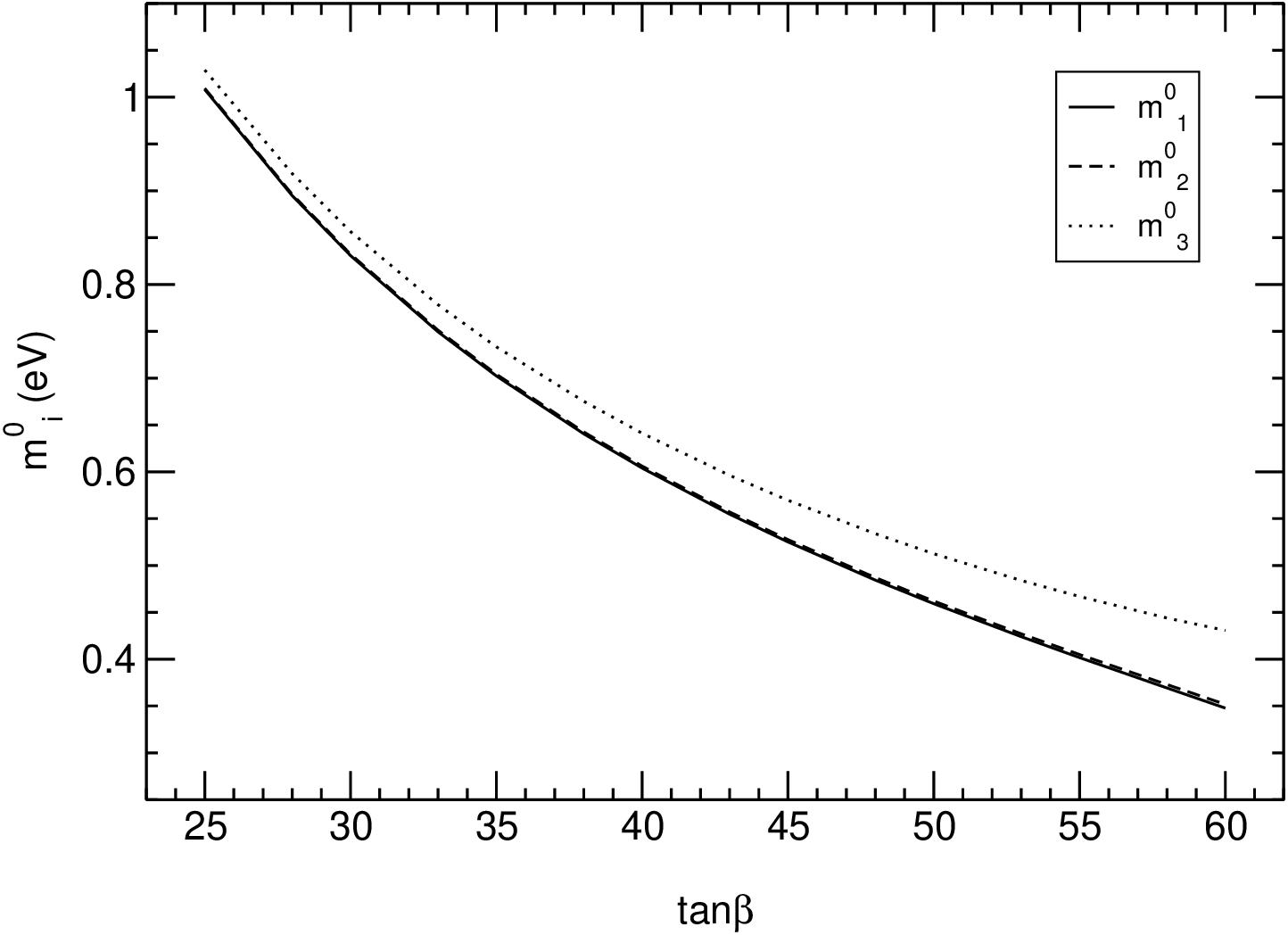}
	\caption{Variation of $\tan\beta$ vs neutrino masses ($m_i^0$) at unification scale. In plotting 
                  this figure, we have taken unification scale $= 2 \times 10^{16}$ GeV and $M_{SUSY} = 2 \times 10^3$ GeV. }
	\label{fig8}
 	\end{center}
\end{figure}

\begin{figure}[h!]
 	\begin{center}
 	 \includegraphics[width = 12.0 cm]{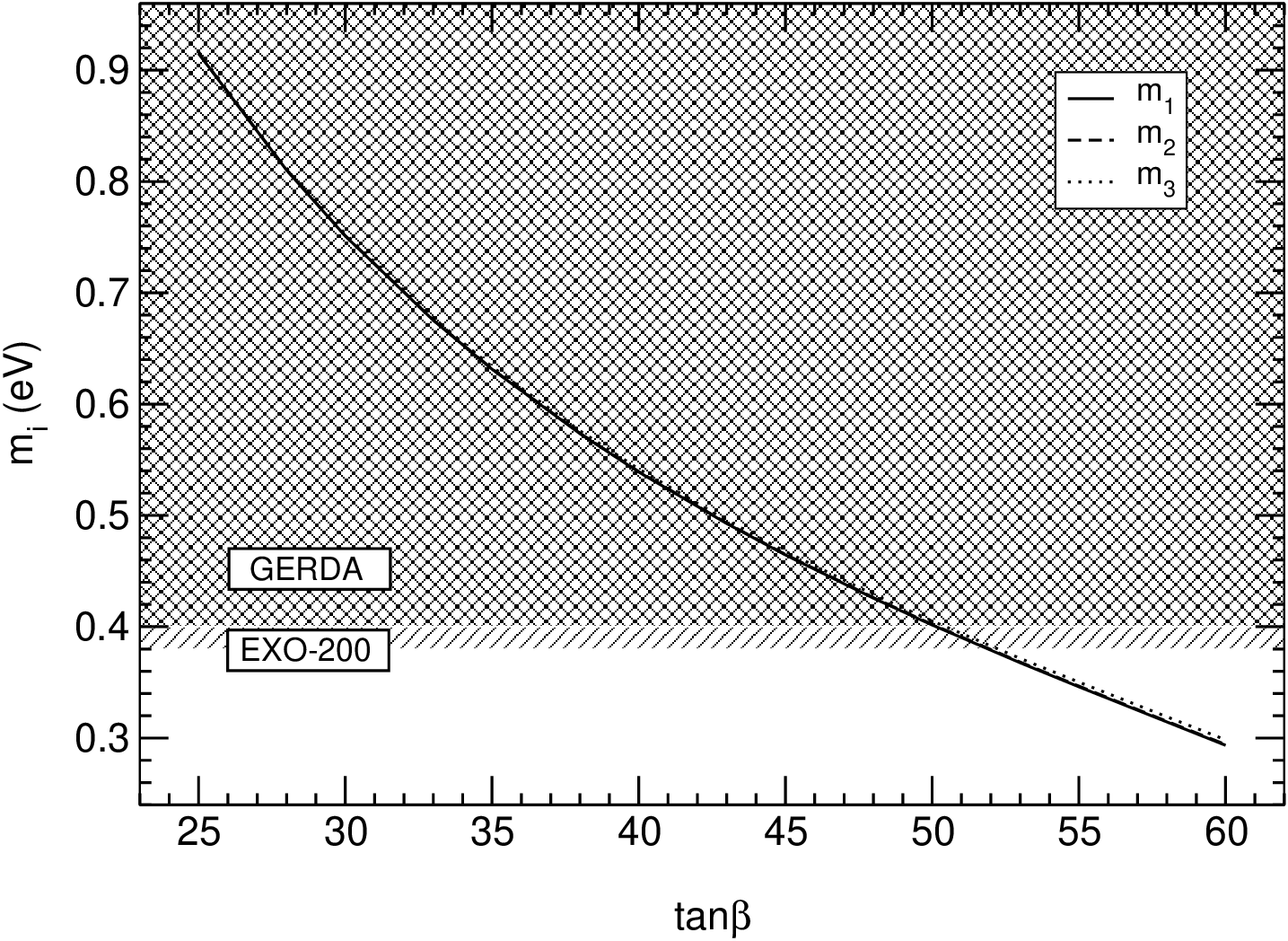}
	\caption{Variation of $\tan\beta$ vs neutrino masses ($m_i$) at $M_Z$. In plotting 
                  this figure, we have taken unification scale $= 2 \times 10^{16}$ GeV and $M_{SUSY} = 2 \times 10^3$ GeV.
                  The shaded regions are excluded by $0\nu\beta\beta$ decay experiments \cite{Agostini:2013mzu,Auger:2012ar}.  }
	\label{fig9}
 	\end{center}
\end{figure} 

It is clear from Figures \ref{fig8} and \ref{fig9} that the mixing angle magnification happens for a wide range of $\tan\beta $.  Although we have not shown this in the figure, the desired angle magnifications can be obtained for values of 
$\tan\beta $ as low as 4 or 5. But, for low $\tan \beta$ the masses of neutrinos become very high at low scale. 
Furthermore, if we take low values of $\tan\beta $, the input neutrino masses, at unification scale, have to be taken more degenerate and the mean mass should also be higher. 
The reason is that with decreasing $\tan\beta $ the factor $(1 + \tan^2\beta)$ becomes small.  
Thus, to achieve desired magnifications at 
$M_Z$ one has to make the input neutrino masses more degenerate to account for the smaller contribution coming from $(1 + \tan^2\beta)$ term. 
At the same time to keep  $\Delta m^2_{\rm{32}}$ within its 3-$\sigma$ range at $M_Z$, one is forced to increase the mean input mass at unification scale.
Since the mean mass of the neutrinos increases with decreasing $\tan\beta $, one can constrain the range of allowed $\tan \beta$ from various experiments, as discussed in Section \ref{subsec44}.


\subsection{Experimental Constraints}
\label{subsec44}


As is clear from previous discussion, the mean mass of neutrinos varies with the variation of the unification scale, SUSY breaking scale and $\tan\beta$. 
Therefore, one can constrain the range of these parameters by using data from various experiments, as discussed below.

(i) {\it Constraints from Tritium Beta Decay:} The present constraints on $\langle m_{ \beta} \rangle$ coming from tritium beta decay are $\langle m_\beta \rangle < 2 $ eV \cite{Beringer:1900zz, Kraus:2004zw, Aseev:2011dq}.
 They give only the upper bound on the masses of the neutrinos
and thus the whole mass range of the Figures \ref{fig5}, \ref{fig7} and \ref{fig9} easily comes under this limit. Hence, the tritium beta decay constraints are relatively weak. They allow much lower values of the unification 
scale, $\tan \beta$ and much higher values of SUSY breaking scale than those plotted in the above figures. However, in future, the KATRIN experiment is expected to probe $\langle m_{ \beta} \rangle$ as low as 0.2 eV \cite{Drexlin:2013lha}
and hence will be able to put much tighter constraints on the allowed range of these parameters.

(ii) {\it Constraints from Neutrinoless Double Beta Decay}: At present, the EXO-200 and GERDA experiments provide the most stringent constraints on $\langle m_{ \beta\beta} \rangle$.
The latest results from phase I of the GERDA experiment have given the upper limit on $\langle m_{ \beta \beta} \rangle$ to be 0.20-0.40 eV \cite{Agostini:2013mzu}, whereas EXO-200 has given an upper limit of 0.14-0.38 eV \cite{Auger:2012ar}. 
This, in turn, puts stringent constraints on the allowed range of various parameters, as given below.  

(a) The lower limit of unification scale is constrained to be around $10^{13}$ GeV by GERDA and around $10^{14}$ GeV by EX0-200 (cf. Figure \ref{fig5}). 

(b) The results from GERDA constrains the highest possible SUSY breaking scale to be around $10^{7}$ GeV, whereas EXO-200 puts a limit of around $10^6$ GeV (cf. Figure \ref{fig7}). 

(c) The lowest possible value of $\tan\beta $ is constrained to be around 50 (cf. Figure \ref{fig9}).

In future, these limits are expected to improve, thus resulting in more tighter constraints e.g. the GERDA phase II is aiming for an increased sensitivity by a factor of about 10 (cf. \cite{Agostini:2013mzu} for details). 
It should be noted that the above constraints are for the case when all the PMNS phases are taken to be zero. These constraints are likely to change in the presence of phases. 
We will analyze them, in detail, in our next work \cite{AGRS}.

(iii) {\it Cosmological Constraints:} The recent result of the Planck collaboration has given constraints on the sum of neutrino masses to be in the range of 0.23 eV [95\%; Planck+WP+highL+BAO]  
to 1.08 eV [95\%; Planck+WP+highL ($A_L$)] depending on values chosen for the priors \cite{Ade:2013zuv}. 
The limit of 1.08 eV implies that the mean neutrino mass has to be around 0.36 eV, thus putting similar constraints to those obtained from  $\langle m_{ \beta \beta} \rangle$. 
The lowest value (i.e. 0.23 eV [95\%; Planck+WP+highL+BAO]) is in tension with our hypothesis. However, as noted by the Planck collaboration itself, the cosmological limits are 
highly dependent on chosen values of priors, so these limits should be taken as indicative and not conclusive.

To conclude, in view of the above experimental constraints, for fixed values of other parameters, 
(1) The unification scale should be taken around $10^{14}$ GeV or above,
(2) The SUSY breaking scale should be taken below $10^6$ GeV and
(3) The $\tan\beta$ should be taken above 50.


\section{Conclusions}
\label{sec6}


We have investigated the implications of High Scale Mixing Unification hypothesis in the wake of new data and experimental constraints. This hypothesis leads to the experimentally observed mixing angles and 
mass square differences at low energy scales ($M_Z$). The small but non-zero value of $\theta_{13}$ is a natural outcome of this hypothesis which has been recently confirmed by various experiments 
\cite{Abe:2011sj,Adamson:2011qu,Abe:2012tg,Ahn:2012nd,An:2012eh}. 
We found that, in absence of phases, for the present 3-$\sigma$ range of $\theta_{13}$ HSMU hypothesis uniquely predicts the value of $\theta_{23}$ to be non-maximal and above $45^\circ$. 
The normal hierarchy and quasi-degeneracy of neutrino masses are essential assumptions to realize HSMU hypothesis. 
We have also analyzed the allowed parameter range for other parameters of our hypothesis  {\it vis-a-vis} various experimental constraints. We found that
(i) the unification scale should be above $ 10^{14}$ GeV, (ii) the SUSY breaking scale should lie below $10^6$ GeV, and (iii) the value of $\tan\beta $ should be taken above 50. 

However, it should be noted that all the above conclusions have been drawn by taking Dirac as well as Majorana phases of PMNS matrix to be zero. 
These conclusions may change in the presence of phases. The detailed implications of these phases are under investigation and will be reported in our future publications \cite{AGRS}. 

Moreover, in a recent analysis of the HSMU hypothesis with Dirac type neutrinos, we have found similar predictions for mixing angles \cite{Abbas:2013uqh}. At the end, we would like to point out that the above 
mentioned two scenarios can be distinguished from each other by the scale of their mean mass (or $\langle m_{ \beta} \rangle$) and $\langle m_{ \beta \beta} \rangle$ measurements.


\begin{acknowledgments}
We would like to thank R. N. Mohapatra, M. K. Parida, S. K. Agarwalla, M. Hirsch and A. Pich for their useful comments and suggestions. 
RS would also like to thank A. Menon, R. Laha and S. Vempati for their valuable comments and suggestions.  
\end{acknowledgments}



\end{document}